\documentclass[singlecolumn,epjc3]{svjour3}
\smartqed  
\RequirePackage{graphicx}
\journalname{Eur. Phys. J. C}

\begin{document}

\title{Generating physically realizable stellar structures via embedding}

\author{S.K. Maurya\thanksref{e1,addr1}
\and M. Govender\thanksref{e2,addr2}.}

\thankstext{e1}{e-mail: sunil@unizwa.edu.om}
\thankstext{e2}{e-mail: megandhreng@dut.ac.za}

\institute{Department of Mathematical and Physical Sciences,
College of Arts and Science, University of Nizwa, Nizwa, Sultanate
of Oman\label{addr1}\and Department of Mathematics, Faculty of Applied Sciences, Durban University of Technology, Durban, South Africa\label{addr2} }

\date{Received: date / Accepted: date}

\maketitle

\begin{abstract}
In this work we present an exact solution of the Einstein-Maxwell field equations describing compact, charged objects within the framework of classical general relativity. Our model is constructed by embedding a four-dimensional spherically symmetric static metric into a five dimensional flat metric. The source term for the matter field is composed of a perfect fluid distribution with charge. We show that our model obeys all the physical requirements and stability conditions necessary for a realistic stellar model. Our theoretical model approximates observations of neutron stars and pulsars to a very good degree of accuracy.
\end{abstract}


\maketitle

\section{Introduction}

Einstein's general theory of relativity has successfully accounted for various observations cosmological scales as well as in astrophysical contexts\cite{tipler1980,shap1983}. The golden age of cosmology has seen the theory fine-tuned to a high degree of accuracy in explaining the Hubble rate, matter content, baryogenesis, nucleosynthesis, as well as the possible origin and subsequent evolution of the Universe. General relativity, as an extension of Newtonian gravity is especially useful in describing compact objects in which the gravitational fields are very strong. Some of these objects include neutron stars, pulsars and black holes where densities are of the order of $10^{14} g.cm^{-3}$ or greater. The first exact solution of the Einstein field equations representing a bounded matter distribution was provided by Schwarzschild in 1916\cite{Sch1916}. This solution described a constant density sphere with the exterior being empty. The constant density Schwarzschild solution was a toy model which cast light on the continuity of the gravitational potentials and the behaviour of the pressure at the surface of the star. However, the interior Schwarzschild solution was noncausal in the sense that it allowed for faster then light propagation velocities within the stellar interior. This prompted the search for physically viable solutions of the Einstein field equations describing realistic stars. A century later and we have thousands of exact solutions of the field equations describing a multitude of stellar objects ranging from perfect fluids, charged bodies, anisotropic matter distributions, higher dimensional stars and exotic matter configurations. Spherical symmetry is the most natural assumption to describe stellar objects. However, there is a wide range of stellar solutions exhibiting departure from sphericity. These solutions include the Kerr metric which describes the exterior gravitational field of a rotating stellar object\cite{kerr1963}. In the limit of vanishing angular momentum, the Kerr solution reduces to the exterior Schwarzschild solution. There have also been numerous attempts at extending the Kerr metric to allow for dissipation and rotation\cite{vaidya1,Car,kramer1}.

In order to generate exact solutions of the Einstein field equations, researchers have employed a wide range of techniques to close this system of highly nonlinear, coupled partial differential equations. In the quest to obtain exact solutions describing static compact objects one imposes (i) symmetry requirements such as spherical symmetry, (ii) an equation of state relating the pressure and energy density of the stellar fluid, (iii) the behaviour of the pressure anisotropy or isotropy, (iv) vanishing of the Weyl stresses, (v) spacetime dimensionality, to name just a few\cite{bowers,sharmaeos,hereos,bharaniso,sharmadark,farookn,govegb}. These assumptions render the problem of finding exact solutions of the field equations mathematically more tractable. There is no guarantee that the resulting stellar model actually describes a physically realizable stellar structure. In the case of nonstatic, radiating stars, various exact solutions are known in the literature ranging from acceleration-free collapse, Weyl-free collapse, vanishing of shear, collapse from/to an initial/final static configuration as well as anisotropic collapse models.

The Randall-Sundrum braneworld scenario has generated an intense interest in higher dimensional gravity and modified theories of gravity\cite{Randal}. Braneworld stars were shown to have nonunique exteriors due to radiative-type stresses arising from 5-dimensional graviton effects emitting from the bulk\cite{germani}. Govender and Dadhich showed that the gravitational collapse of a star on the brane is accompanied by Weyl radiation\cite{govbrane}. They concluded that a collapsing sphere on the brane is enveloped by the brane generalised Vaidya solution which is in turn matched to the Reissner-Nordstrom metric. The mediation of the Vaidya envelope is a unique feature of the braneworld collapse which is not present in standard 4-d Einstein gravity. A recent model by Banerjee et al. showed that Weyl stresses lead naturally to anisotropic pressures within the core of a braneworld gravastar\cite{bangrav}. In their model the Mazur and Mottola gravastar picture \cite{mot} is considered within the Randall-Sundrum II type braneworld scenario.  Recently, Dadhich and coworkers demonstrated the universality of the constant density Schwarzschild solution in general Einstein-Lovelock gravity and the universality of the isothermal sphere for pure Lovelock gravity when $d \geq 2N + 2$. In a recent paper by Chakrabory and Dadhich, they ask a pertinent question: "Do we really live in four dimensions or higher?" This question arises from the fact that while gravity is free to propagate in higher dimensions while all other matter fields are confined to 4-dimensions, gravity cannot distinguish between 4-d Einstein or in particular, 7-d pure Gauss-Bonnet dynamics\cite{dad2}.

The idea of embedding a purely gravitational field represented by a 4-dimensional Riemannian metric into a flat space of higher dimensions has resurrected interest in so-called class one spacetimes. Karmarkar derived the necessary condition for a general spherically symmetric metric to be of class one\cite{kar48}. In general, if the lowest number of dimensions of flat space in which a Riemannian space of dimension $n$ can be embedded in $n + p$, then the Riemannian space is referred to as class $p$. Class one spacetimes have been successfully utilised to model compact objects such as strange star candidates, neutron stars and pulsars\cite{k1,k2,k3,k4,k5,k6,k7}. These theoretical models accurately predict and agree with observations regarding the masses, radii, compactness and densities of these objects within experimental error. On the other hand Momeni et al. \cite{Momeni1,Momeni2,Momeni3} have obtained the realistic compact objects for Tolman-Oppenheimer-Volkoff equations in $f(R)$ gravity in different context.

In this work we use the condition arising from embedding a 4-d spherically symmetric static metric in Schwarzschild coordinates into a 5-d flat spacetime to model a charged compact object. By choosing one of the metric potentials on physical grounds, the embedding condition gives us the second metric potential which then completely describes the gravitational behaviour of the model. This paper is structured as follows: In Section two we introduce the 4-d Einstein spacetime and provide the necessary and sufficient condition for embedding this spacetime into a 5-d flat spacetime. The Einstein-Maxwell field equations describing the gravitational behaviour of our stellar model are presented in Section three. In Section four we derive an exact solution of the Einstein-Maxwell equations describing a charged, static sphere by making use of the embedding condition derived in the previous section. The boundary conditions required for the smooth matching of the interior of the star to the vacuum Schwarzschild exterior solution is given in Section five. The physical viability of our model is considered in Section six. We conclude with a discussion in Section seven.

\section{Class one condition for Spherical symmetric metric:}

The spherically symmetric line element in Schwarzschild co-ordinates $(x^{i})=(t,r,\theta,\phi)$ is given as:
\begin{equation}
ds^{2}=e^{\nu(r)}dt^{2}-e^{\lambda(r)}dr^{2}-r^{2}\left(d\theta^{2}+\sin^{2}\theta d\phi^{2} \right)\label{eq1}
\end{equation}
where $\lambda$ and $\nu$ are the functions of the radial coordinate $r$.\\

To determine the class one condition of the metric above (\ref{eq1}), we suppose that the 5-dimensional metric is flat

\begin{equation}
ds^{2}=-\left(dz^1\right)^2-\left(dz^2\right)^2-\left(dz^3\right)^2-\left(dz^4\right)^2+\left(dz^5\right)^2,\label{eq2}
\end{equation}

where \,\, $z^1=r\,sin\theta\,cos\phi$, \,\, $z^2=r\,sin\theta\,sin\phi$, \,\,\, $z^3=r\,cos\theta$, \\

 $z^4=\sqrt{K}\,e^{\frac{\nu}{2}}\,cosh{\frac{t}{\sqrt{K}}}$,\,\,\,$z^5=\sqrt{K}\,e^{\frac{\nu}{2}}\,sinh{\frac{t}{\sqrt{K}}}$,.\\
and $K$ is a positive constant.
On inserting the components $ z^1,z^2, z^3, z^4 $ and $z^5$ into the metric (\ref{eq2}), we obtain
\begin{equation}
ds^{2}=-\left(\,1+\frac{K\,e^{\nu}}{4}\,{\nu'}^2\,\right)\,dr^{2}-r^{2}\left(d\theta^{2}+\sin^{2}\theta d\phi^{2} \right)+e^{\nu(r)}dt^{2},\label{eq3}
\end{equation}

Comparing the line element (\ref{eq3}) with the line element (\ref{eq1}) we get

\begin{equation}
e^{\lambda}=\left(\,1+\frac{K\,e^{\nu}}{4}\,{\nu'}^2\,\right),\label{eq4}
\end{equation}

 The condition (\ref{eq4}) implies that the class of metric is one because we have embedded 4-dimensional space time into 5-dimensional flat space time. We should point out that (\ref{eq4}) is equivalent to the condition derived by Karmarkar in terms of the Riemann tensor components
 \begin{equation}
 {\cal R}_{1414}{\cal{R}}_{2323} = {\cal{R}}_{1212}{\cal{R}}_{3434} + {\cal{R}}_{1224}{\cal{R}}_{1334}
 \end{equation}
 where ${\cal{R}}_{2323}\neq 0$.

\section{Einstein-Maxwell field equations}

The Einstein-Maxwell field equations can be written as

\begin{equation}
8\pi\,\left(\,T_{\nu}^{\mu}+E_{\nu}^{\mu}\,\right)= R_{\nu}^{\mu}-\frac{1}{2}\,R\,g_{\nu}^{\mu},\label{eq5}
\end{equation}
Here we assume that the matter is a perfect fluid within the star, then $T_{\nu}^{\mu}$ and $E_{\nu}^{\mu}$ are the corresponding energy-momentum tensor and electromagnetic field tensor, respectively defined by
\begin{eqnarray}
T^\nu_\mu &=& (\rho + p)v^\nu v_\mu - p\delta^\nu_\mu \\
E^\nu_\mu &=& \frac{1}{4\pi}\left(-F^{\nu\gamma}F_{\mu\gamma} + \frac{1}{4}\delta^\nu_\mu F^{\gamma\mu}F_{\gamma\mu}\right),
\end{eqnarray}
where $\rho$ is the energy density, $p$ is the isotropic pressure and $v^\nu$ is the fluid four-velocity given as $e^{-\nu(r)/2}v^\nu=\delta^\nu_\mu$.
We are using geometrized units and thus take $\kappa=8\pi$ and $G=c=1$. The components of $T^\nu_\mu$ and $E^\nu_\mu$ are as follows:
${T^1}_1=-p,\, {T^2}_2={T^3}_3=-p,\, {T^4}_4=\rho$ and ${E^1}_1=-{E^2}_2=-{E^3}_3={E^4}_4=\frac{1}{8\,\pi}\,\frac{q^2}{r^4}$.\\

For the spherically symmetric metric Eq.(\ref{eq1}), the Einstein-Maxwell field equations (\ref{eq5}) are (\cite{Dionysiou}):

\begin{equation}
\frac{e^{-\lambda}-1}{r^{2}}+\frac{e^{-\lambda}\nu'}{r}=8\pi\, p-\frac{q^2}{r^4} \label{eq6}
\end{equation}

\begin{equation}	
e^{-\lambda}\left(\frac{\nu''}{2}+\frac{\nu'^{2}}{4}-\frac{\nu'\lambda'}{4}+\frac{\nu'-\lambda'}{2r} \right)=8\pi\, p + \frac{q^2}{r^4}.\label{eq7}
\end{equation}

\begin{equation}
\frac{1-e^{-\lambda}}{r^{2}}+\frac{e^{-\lambda}\lambda'}{r}=8\pi\,\rho + \frac{q^2}{r^4},\label{eq8a}	
\end{equation}

If we now demand that the radial and transverse stresses are equal at each interior point of the stellar configuration we obtain from equating  Eqs. (\ref{eq6})  and (\ref{eq7})

\begin{equation}
\frac{2\,q^2}{r^4}=	e^{-\lambda}\,\left[\frac{2\,\nu''-\nu'\,\lambda'+\nu'^{2}}{4}-\frac{\lambda'+\nu'}{2r} \right]-\frac{e^{-\lambda}-1}{r^2}\,\label{eq8b}
\end{equation}
known as the condition of pressure isotropy. We note that the Einstein-Maxwell equations (\ref{eq6}) -- (\ref{eq8a}) can be viewed as describing an perfect fluid with anisotropic pressure. Eqn. (\ref{eq8b}) can be used as a definition for the electric field intensity. Alternatively, if we specify the nature of the electric field intensity then Eqn. (\ref{eq8b}) gives a relation between $\nu$ and $\lambda$. This is a common approach in solving the Einstein-Maxwell system. In our approach we will utilise the embedding condition given in Eqn. (\ref{eq4}) to obtain an exact solution of the Einstein-Maxwell field equations.

\noindent However if $m(r)$ is the mass function for electrically charged compact star model then it can be defined in terms of metric function $e^{\lambda}$
 and electric charge $q$ as

\begin{equation}
m(r)=\frac{r}{2}\,\left[\,1-e^{-\lambda(r)}+\frac{q^2}{r^2} \,\right]  \label{mass1}
\end{equation}

\section{New class of general solutions for a charged compact star:}

We note that Eqn. (\ref{eq4}) relates the metric functions $\nu$ and $\lambda$ thus reducing the task of finding exact solutions to a single-generating function.
Now to determine the mass function $m(r)$ and electric charge $q$, we assume the following form for the metric function $e^{\nu}$:
\begin{equation}
e^{\nu}=B\,(1-Ar^2)^n,\label{eq10a}
\end{equation}

where, $A$ and $B$ are positive constants and $n \le -1$. This form of the metric function is well-motivated and has been utilised by Maurya et al.\cite{maurya11} to model charged compact stars arising from the Karmarkar condition. In these models they took $n > 2$. The parameter $n$ acts as a 'switch' and characterises various well-known models available in the literature. It is clear from Eqn. (\ref{eq10a}) that $n = 0$ renders the spacetime flat which is meaningless in the present context of this paper. It was first pointed out by Tikekar and more recently by Maurya et al.\cite{maurya} that the Karmarkar condition together with isotropic pressure (in the case of neutral fluids) admits two solutions: the Schwarzschild interior solution and the Kohler-Chao-Tikekar solution\cite{kc,tik}. The Schwarzschild solution is conformally flat, ie., the Weyl tensor vanishes at each interior point of the sphere. The Kohler-Chao-Tikekar solution is not conformally flat and furthermore represents a cosmological solution. This is to say that there is no finite radius for which the pressure vanishes in the Kohler-Chao-Tikekar solution. We regain the Kohlar-Chao-Tikekar solution when we set $n = 1$ in Eqn. (\ref{eq10a}). Furthermore, we observe from Table 3 that the product $nA$ is approximately constant for large $n$. As pointed out here that as $n \rightarrow -\infty$ the metric function $\nu = Cr^2 + \ln{B}$ where we have defined here $C = - nA$. This form of the metric function $\nu$ has been already used to construct electromagnetic mass (EMMM) models by Maurya et al.\cite{maurya}. These models have the peculiar feature of vanishing electromagnetic field, mass, pressure and density when the parameter $n = 0$. In addition, the fluid obeys an equation of state of the form $p + \rho = 0$ implying that the pressure within the bounded configuration is negative. In this study we will consider solutions for $n < 0$.
We should point out that the solution describes a physical viable compact star when $n \geq -2.7$. For $n < -2.7$, causality is violated within the stellar fluid as the sound speed exceeds unity. We have started our physical analysis with $n =-6.5$ since there are no physically realizable stars between n=-2.7 to -6.5 as observed by Gangopadhyay et al.\cite{GA} In the limiting case n= -2.7 one expects low mass stars.

\begin{figure}[!htbp]\centering
    \includegraphics[width=5.5cm]{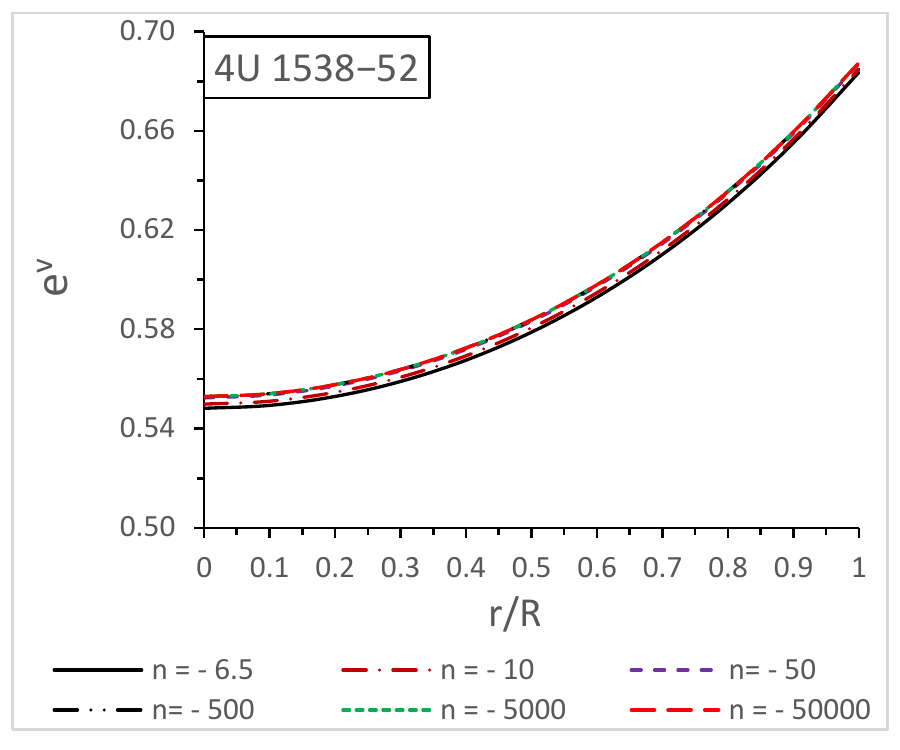}
\caption{Variation of metric function $e^{\nu}$ with the radial coordinate ($r/R$) for $4U 1538-52$ with mass $(M)= 0.87 M_\odot$ and radius $(R)=7.866Km $ (Table \ref{TableA1}). For plotting of this figure, The values of constants for different $n$ are as follows:(i) A=5.3980 $\times 10^{-4} $,  B=0.548101, D=10.38552,  K=8.30822$\times 10^{2}$   for $n=-6.5$, (ii) A=3.5121$\times 10^{-4} $ , B=0.549818, D=16.26344, K=8.422215$\times 10^{2}$ for $n=-10$, (iii) A=7.0350$\times 10^{-5}$, B=0.552284, D=83.46073, K=8.592418 $\times 10^{2}$ for $n=-50$, (iv) A=7.03753$\times 10^{-6}$, B=0.552841, D=8.39474$\times 10^{2}$, K=8.630715 $\times 10^{2}$ for $n=-500$, (v) A=7.03753$\times 10^{-7}$, B=0.552899, D=8.39963$\times 10^{3}$, K=8.634831 $\times 10^{2}$ for $n=-5000$, (vi) A=7.03753$\times 10^{-8}$, B=0.552905, D=8.40010$\times 10^{4}$, K=8.635231 $\times 10^{2}$ for $n=-50000$ (Table\ref{TableA3}).}
    \label{Fig1}
\end{figure}

Now by plugging our choice of $e^{\nu}$ from Eq.(\ref{eq10a}) into Eq.(\ref{eq4}), we obtain

\begin{equation}
e^{\lambda}=[1+D\,Ar^2\,(1-Ar^2)^{n-2}],\label{eq10b}
\end{equation}

where $D=n^2\,A\,B\, K $.

\begin{figure}[!htbp]\centering
    \includegraphics[width=5.5cm]{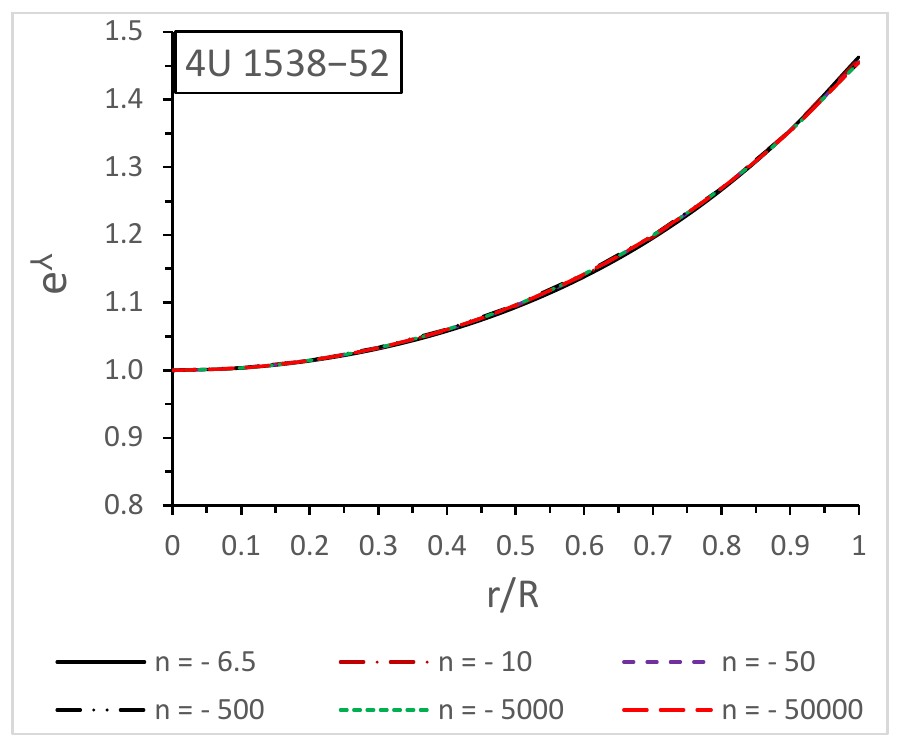}
\caption{Variation of metric function $e^{\lambda}$ with the radial coordinate ($r/R$) for $4U 1538-52$ with mass $(M)= 0.87 M_\odot$ and radius $(R)=7.866Km $ (Table \ref{TableA1}). For plotting of this figure, we have employed same values of the constant as used in Fig.1.  The corresponding numerical values can be seen from Table\ref{TableA3}.}
    \label{Fig2}
\end{figure}

Then on inserting $e^{\nu}$ and $e^{\lambda}$ from Eqs.(\ref{eq10a}) and (\ref{eq10b}) respectively into Eqs.(\ref{eq8b}) and (\ref{mass1}), we get:

\begin{equation}
\frac{2q^2}{r^4}=Ar^2 \left[\frac{n^2\,{\psi^2}-2\,n\,\psi\,({\psi}-D\,{\psi^n})+D\,{\psi^n}\,(-2{\psi}+D\,{\psi^n})}{({\psi^2} + D\,Ar^2\,{\psi^n})^2} \right] \label{eq11}
\end{equation}

\begin{equation}
m(r)=\frac{A^2r^5\,[3D^2\psi^{2n}+n(n-2)\,\psi^2]+2\,D\,Ar^3\,\psi^{n+1}[1+(n-2)\,Ar^2]}{4 \left[{\psi^2} + D\,Ar^2\,{\psi^n}\right]^2}  \label{eq12}
\end{equation}

where, $\psi=(1-Ar^2)$,\,\,

\begin{figure}[!htbp]\centering
    \includegraphics[width=5.5cm]{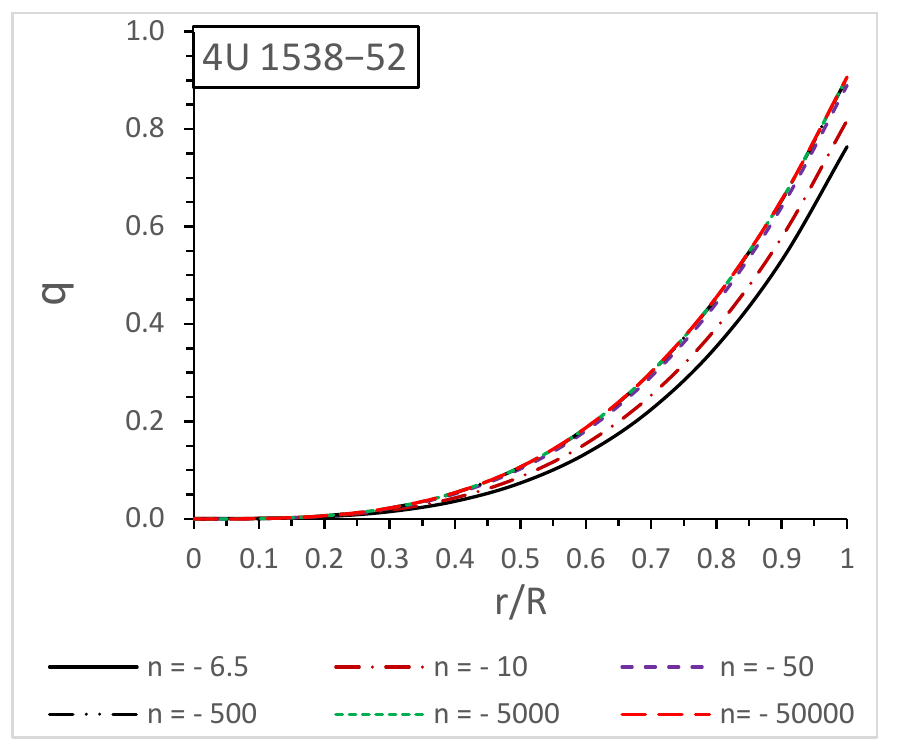}\includegraphics[width=5.5cm]{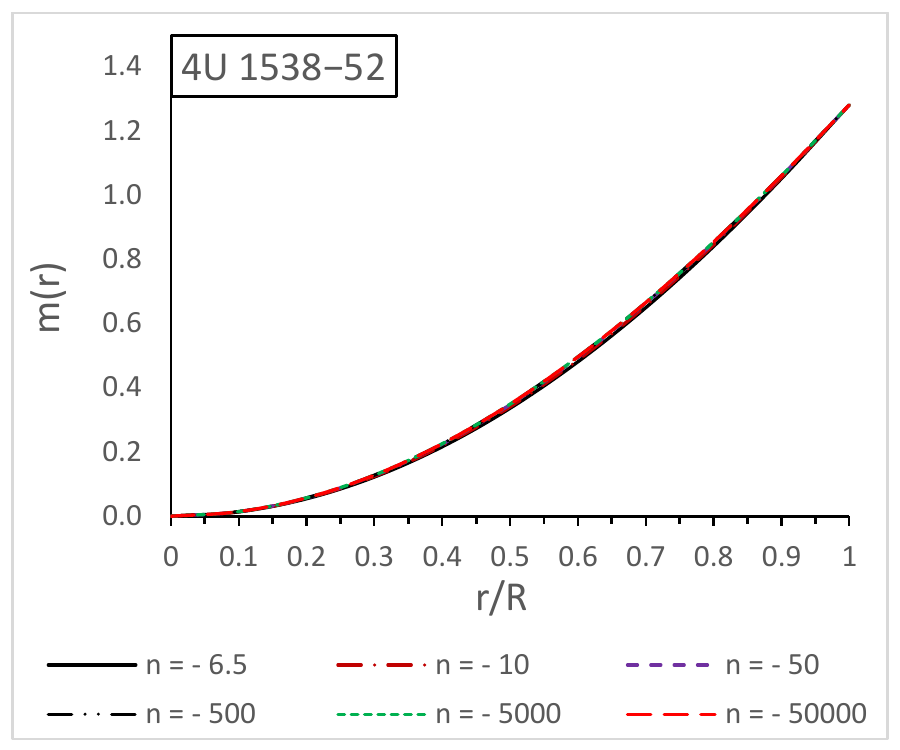}
\caption{Variation of electric charge, $q$, (left panel) and mass function, $m(r)$, (right panel) with the radial coordinate ($r/R$) for $4U 1538-52$ with mass $(M)= 0.87 M_\odot$ and radius $(R)=7.866Km $ (Table \ref{TableA1}). For plotting of this figure, we have employed same values of the constant as used in Figs.1 and 2. The corresponding numerical values can be seen from Table\ref{TableA3}.}
    \label{Fig3}
\end{figure}

The expressions for the pressure and energy density are determined from Eqs(\ref{eq6}) and (\ref{eq8a}) respectively and can be written as

\begin{equation}
\frac{8\pi\,p}{A}=\frac{n^2\,Ar^2\,\psi^{2}-D\,\psi^{n}\,[2\psi+D\,Ar^2\,\psi^{n}]-2\,n\,\psi\,[(2-Ar^2)\,\psi+D\,Ar^2\,\psi^n]}{2\left[{\psi^2} + D\,Ar^2\,{\psi^n}\right]^2} \label{eq13}
\end{equation}

\begin{figure}[!htbp]\centering
    \includegraphics[width=5.5cm]{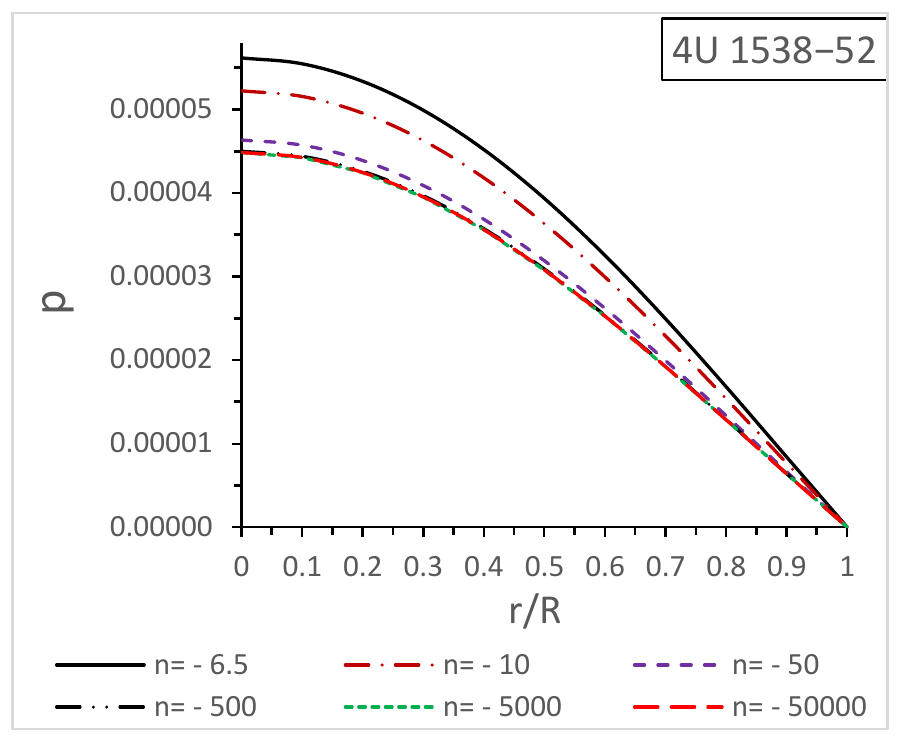}
\caption{Variation of pressure $p$ with the radial coordinate ($r/R$) for $4U 1538-52$ with mass $(M)= 0.87 M_\odot$ and radius $(R)=7.866Km $ (Table \ref{TableA1}). For plotting of this figure, the values of constants for different $n$ are as follows: (i) A=5.3980 $\times 10^{-4} $,  B=0.548101, D=10.38552,  K=8.30822$\times 10^{2}$   for $n=-6.5$, (ii) A=3.5121$\times 10^{-4} $ , B=0.549818, D=16.26344, K=8.422215$\times 10^{2}$ for $n=-10$, (iii) A=7.0350$\times 10^{-5}$, B=0.552284, D=83.46073, K=8.592418 $\times 10^{2}$ for $n=-50$, (iv) A=7.03753$\times 10^{-6}$, B=0.552841, D=8.39474$\times 10^{2}$, K=8.630715 $\times 10^{2}$ for $n=-500$, (v) A=7.03753$\times 10^{-7}$, B=0.552899, D=8.39963$\times 10^{3}$, K=8.634831 $\times 10^{2}$ for $n=-5000$, (vi) A=7.03753$\times 10^{-8}$, B=0.552905, D=8.40010$\times 10^{4}$, K=8.635231 $\times 10^{2}$ for $n=-50000$ (Table\ref{TableA3}).}
    \label{Fig4}
\end{figure}

\begin{equation}
\frac{8\pi\,\rho}{A}=\frac{D^2\,Ar^2\psi^{2n}-n(n-2)\,Ar^2\,\psi^{2}-2D\,\psi^{n+1}\,[-3+(3n-2)\,Ar^2]}{2\left[{\psi^2} + D\,Ar^2\,{\psi^n}\right]^2} \label{eq14}
\end{equation}

\begin{figure}[!htbp]\centering
    \includegraphics[width=5.5cm]{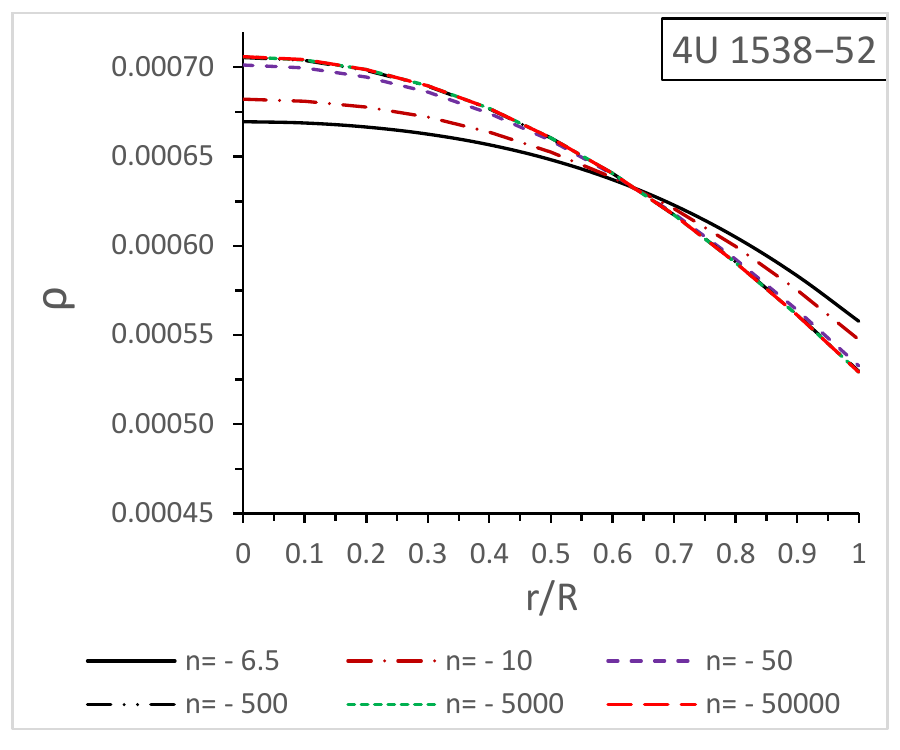}
\caption{Variation of density $\rho$ with the radial coordinate ($r/R$) for $4U 1538-52$ with mass $(M)= 0.87 M_\odot$ and radius $(R)=7.866Km $ (Table \ref{TableA1}). For plotting of this figure, we have employed same values of the constant as used in Figs.3. The corresponding numerical values can be seen from Table\ref{TableA3}.}
    \label{Fig5}
\end{figure}

\section{Boundary conditions}

In order to fix the constants appearing in our solution the following conditions must be satisfied:
(i) The interior metric must join smoothly with the exterior Reissner-No$\ddot{r}$dstrom metric at the boundary of charged compact star ($r=R$). The Reissner-No$\ddot{r}$dstrom metric take takes the form
\begin{equation}
    \label{eq15}
ds^{2} =\left(1-\frac{2M}{r}+\frac{q^2}{r^2}\right)\, dt^{2} -\left(1-\frac{2M}{r}+\frac{q^2}{r^2}
\right)^{-1} dr^{2}-r^{2} (d\theta ^{2}
+\sin ^{2} \theta \, d\phi ^{2} ),
\end{equation}

where $M$ is a constant representing the total mass of the charged compact star.

(ii) The radial pressure $p_{r}$ must vanish at the boundary ($r = R$) of the star ( i.e. the continuity of  $\frac{\partial g_{tt}}{\partial r}$ across the boundary of the star) ~\cite{Misner1964}, which is known as the second fundamental form.\\

Vanishing of the radial pressure at boundary $p_r(R)$  =0 yields:
\begin{equation}
D=\frac{-2\,\Psi^{n+1}\,(1+n\,AR^2)+\sqrt{4\,(1+n\,AR^2)^2\,\Psi^{2n+2}+4\,AR^2\,\Psi^{2n}\,\Phi(R)}}{2\,AR^2\,(1-AR^2)^{2n}} \label{eq16}
\end{equation}

where we have defined

\begin{eqnarray}
\Psi &=&(1-AR^2),\\
\Phi(R)&=&[-4n+10n\,AR^2+n^2\,AR^2-2n\,A^2R^4\,(4+n)+n\,A^3R^6\,(2+n)]. \end{eqnarray}

The constant $B$ can be determined by using the condition $e^{\nu(R)}=e^{-\lambda(R)}$, which yields:

\begin{equation}
B=\frac{1}{(1-AR^2)^n\,[1+D\,AR^2\,(1-AR^2)^{n-2}]}   \label{eq17}
\end{equation}

The condition $e^{-\lambda(R)}=1-\frac{2M}{R}+\frac{Q^2}{R^2}$ gives the total mass of the charged compact star as:
\begin{equation}
\frac{M}{R}=\frac{A^2R^4\,[3D^2\Psi^{2n}+n(n-2)\,\Psi^2]+2\,D\,AR^2\,\Psi^{n+1}[1+(n-2)\,AR^2]}{4 \left[{\Psi^2} + D\,AR^2\,{\Psi^n}\right]^2}    \label{eq18}
\end{equation}

By using the density of the star at surface, the value of constant $A$ can be determined from the expression:
\begin{equation}
A=\frac{16\pi\,\rho_s[\Psi^2+D\,AR^2\,\Psi^n]^2}{D^2\,AR^2\Psi^{2n}-n(n-2)\,AR^2\,\Psi^{2}-2D\,\Psi^{n+1}\,[-3+(3n-2)\,AR^2]}   \label{eq19}
\end{equation}

The expressions for the pressure gradient and density gradient, respectively are:
 \begin{equation}
\frac{8\pi\,dp}{dr}=\frac{2\,A^2\,r\,[2n^3\,D\,\psi^{n+1}\,A^2r^4-n^2\,\phi_1(r) +\phi_2(r)+\phi_3(r)]}{2\,\left[{\psi^2} + D\,Ar^2\,{\psi^n}\right]^2}\label{eq20}
\end{equation}

 \begin{equation}
\frac{8\pi\,d\rho}{dr}=\frac{2\,A^2\,r\,[-2D\,n^3\,\psi^{n+1}\,A^2r^4+n^2\phi_4(r)+\phi_5(r)+2n\,\phi_6(r)]}{2\,\left[{\psi^2} + D\,Ar^2\,{\psi^n}\right]^3}\label{eq21}
\end{equation}

where,\\
 $ \phi_1(r)=[-1+Ar^2(2+7D\psi^{n})-2D\psi^{n}\,A^2r^4\,(4-D\,\psi^{n})-(2-D\,\psi^{n})A^3r^6+A^4r^8]$,\\

$\phi_2(r)=2\,n\,\psi\,[(Ar^2-3)\,\psi^{2}+D^2\,Ar^2\,\psi^{2n}+D\,\psi^{n}\,(4-3Ar^2+A^2r^4)]$,\\

$\phi_3(r)=D\,\psi^{n}\,[-6\,\psi^{2}+D^2\,\psi^{2n}\,Ar^2+D\,\psi^{n}\,(3-4Ar^2+3A^2r^4)]$,\\

$ \phi_4(r)=[-1+(2+7D\,\psi^{n})Ar^2-2\,D\,\psi^{n}\,A^2r^4\,(4+3D\,\psi^{n})-(2-D\,\psi^{n})A^3r^6+A^4r^8]$,\\

$\phi_5(r)=-D\,\psi^{n}\,[D^2\,\psi^{2n}\,Ar^2-2\,\psi^{2}\,(11+4\,Ar^2)+D\psi^{n}\,(11-4Ar^2+3\,A^2r^4)]$,\\

$\phi_6(r)=[\psi^{3}(1+Ar^2)+D^2\,\psi^{2n}\,Ar^2\,(5+3Ar^2)-D\,\psi^{n}\,(6-3Ar^2-10A^2r^2+7A^3r^6)]$.

\section{Physical properties of the solution:}

\subsection{Regularity }

\noindent (i) Metric functions at the centre, $r=0$: we observe from Eqs. (\ref{eq10a}) and
(\ref{eq10b}) that the metric functions at the centre $r=0$ assume the values $e^{\nu(0)}=B$  and $e^{\lambda(0)}=1$. This shows that
metric functions are free from singularity and positive at the centre (since $B$ is positive).
Also, both metric functions $e^{\nu}$ and $e^{\lambda}$ are monotonically increasing function of $r$ (Fig. 1 \& 2).\\

\noindent (ii) Pressure at the centre $r=0$: From Eq.(\ref{eq13}), we obtain the pressure $p$ at centre $r=0$ as
 $p_0=-A\,(D+2n)/8\,\pi$. Since $A$ and $D$ are positive, it follows that the central pressure is positive provided that $D < - 2n$.\\

\noindent (iii) Matter density at the centre $r=0$: We require that the matter density be positive at central point of the star.
Observation of Eq.(\ref{eq13}) gives us $\rho_{0}=(3\,A\,D/8\,\pi)$. Since $A$ and $D(=A\,B\,n^{2}\,K)$ are
positive due to positivity of $A$,\, $B$,\,$n^{2}$ and \,$K$. This implies that the central density $\rho_c$ is positive.

\subsection{Causality}

Causality requires that the speed of sound be less than the speed of light within the stellar interior. The speed of sound for the charged fluid sphere should be monotonically decreasing from centre to the boundary of the star ($v=\sqrt{dp/d\rho} < 1$). It is clear from Fig. (6) that speed of sound is monotonically decreasing away from the centre and less than $1$. This implies that our fluid model fulfills causality requirements.

\begin{figure}[!htbp]\centering
    \includegraphics[width=5.5cm]{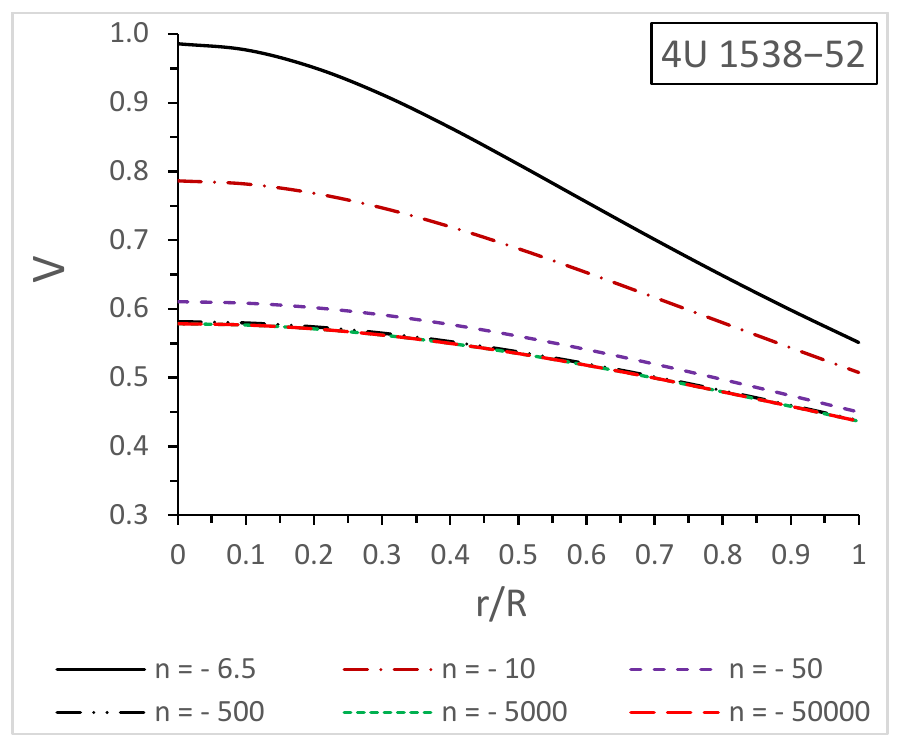}
\caption{Variation of sound velocity $V$ with the radial coordinate ($r/R$) for $4U 1538-52$ with mass $(M)= 0.87 M_\odot$ and radius $(R)=7.866Km $ (Table \ref{TableA1}). For plotting of this figure, The values of constants for different $n$ are as follows: (i) A=5.3980 $\times 10^{-4} $,  B=0.548101, D=10.38552,  K=8.30822$\times 10^{2}$   for $n=-6.5$, (ii) A=3.5121$\times 10^{-4} $ , B=0.549818, D=16.26344, K=8.422215$\times 10^{2}$ for $n=-10$, (iii) A=7.0350$\times 10^{-5}$, B=0.552284, D=83.46073, K=8.592418 $\times 10^{2}$ for $n=-50$, (iv) A=7.03753$\times 10^{-6}$, B=0.552841, D=8.39474$\times 10^{2}$, K=8.630715 $\times 10^{2}$ for $n=-500$, (v) A=7.03753$\times 10^{-7}$, B=0.552899, D=8.39963$\times 10^{3}$, K=8.634831 $\times 10^{2}$ for $n=-5000$, (vi) A=7.03753$\times 10^{-8}$, B=0.552905, D=8.40010$\times 10^{4}$, K=8.635231 $\times 10^{2}$ for $n=-50000$ (Table \ref{TableA3}).}
    \label{Fig6}
\end{figure}

\subsection{Energy conditions}
The charged fluid sphere should satisfy the following three energy conditions, viz., (i)null energy condition (NEC), (ii) weak energy
condition (WEC) and (iii) strong energy condition (SEC).
For satisfying the above energy conditions, the following inequalities must be hold simultaneously inside the charged fluid sphere:

\begin{equation}
NEC: \rho+\frac{E^2}{8\pi}\geq 0,\label{eq22}
\end{equation}

\begin{equation}
WEC: \rho+p \geq  0 \label{eq23}
\end{equation}

\begin{equation}
SEC: \rho+3p-\frac{E^2}{4\pi} \geq  0.\label{eq24}
\end{equation}

It is clear from Fig. (7) that all three energy conditions are satisfied at each interior point of the configuration.

\begin{figure}[!htbp]\centering
    \includegraphics[width=5.5cm]{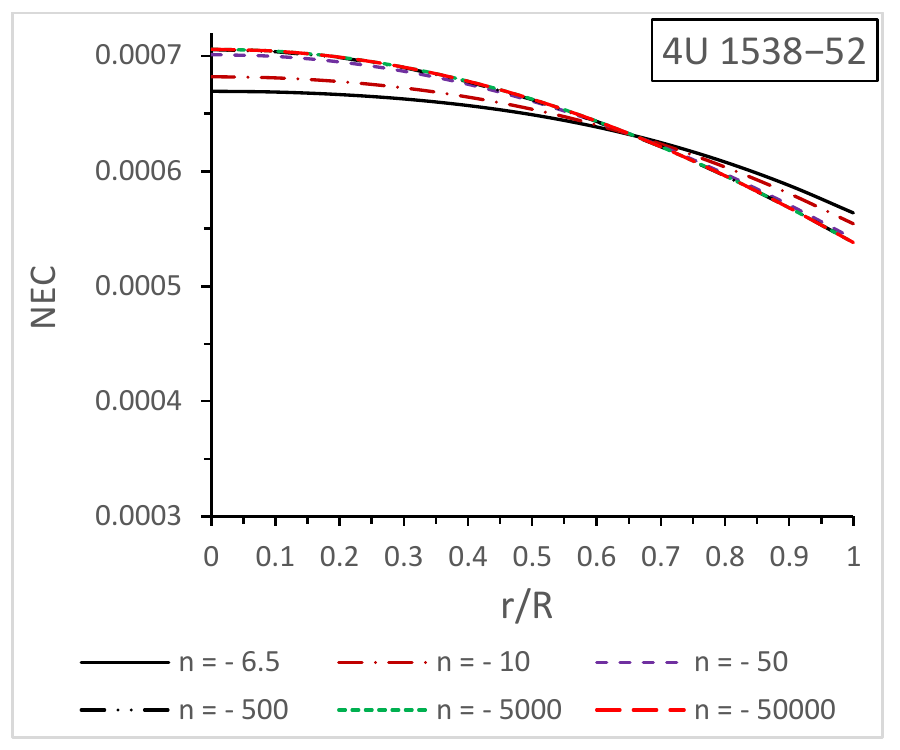} \includegraphics[width=5.5cm]{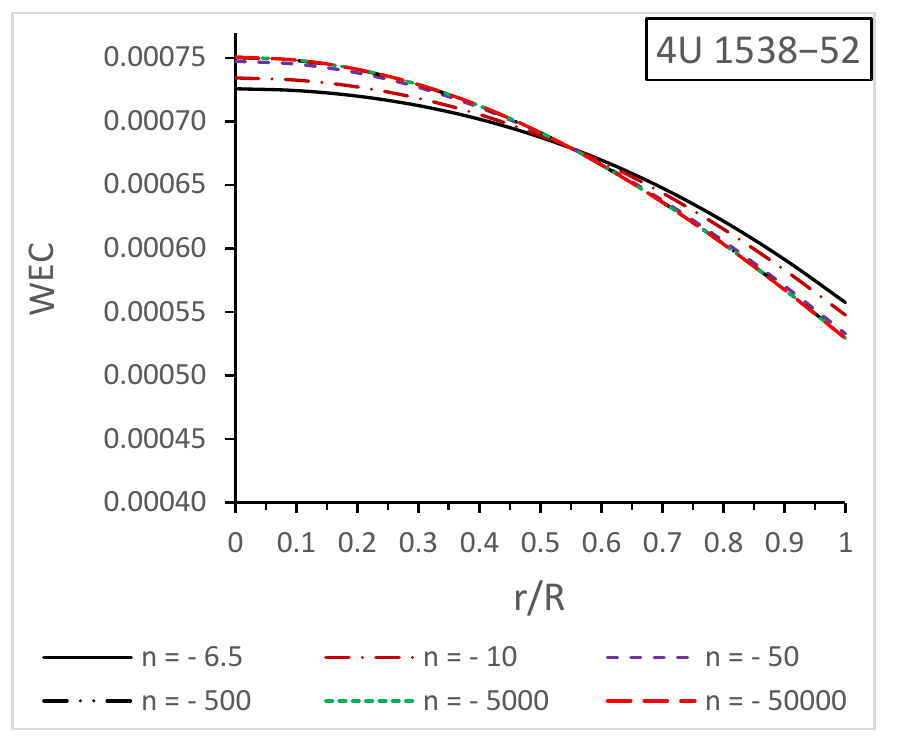} \includegraphics[width=5.5cm]{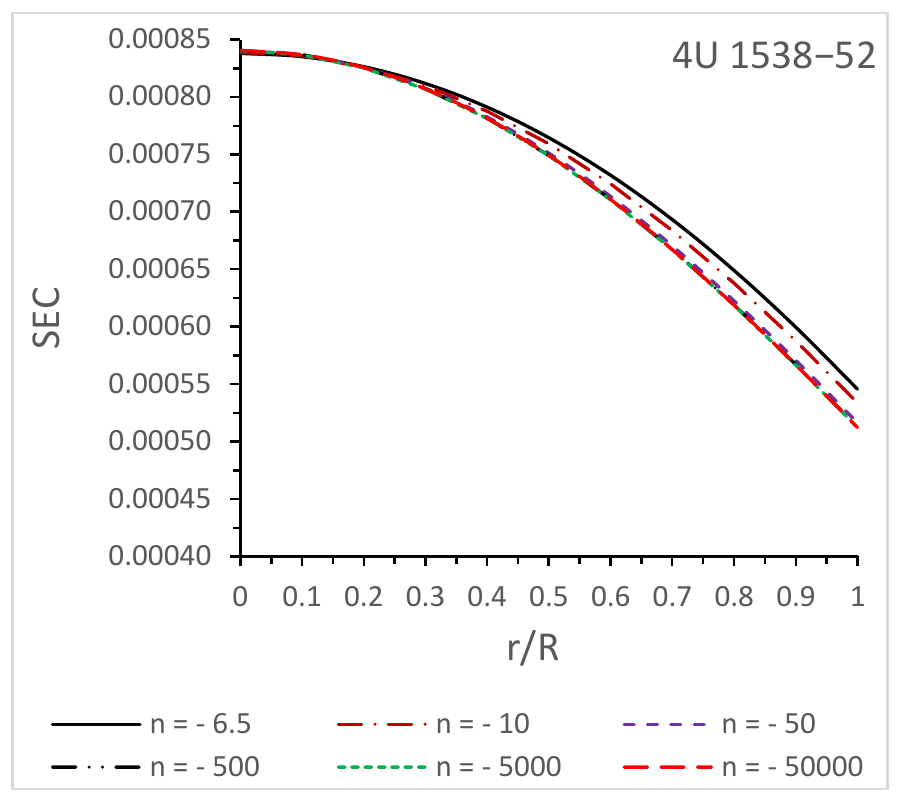}
\caption{Variation of energy conditions $NEC$ ( Top left), $WEC$ (Top right) and $SEC$ (bottom) with the radial coordinate ($r/R$) for $4U 1538-52$ with mass $(M)= 0.87 M_\odot$ and radius $(R)=7.866Km $ (Table \ref{TableA1}). For plotting of this figure, we have employed same values of the constant as used in Figs.6. The corresponding numerical values can be seen from Table\ref{TableA3}.}
    \label{Fig7}
\end{figure}

\subsubsection{Equilibrium condition}

The Tolman-Oppenheimer-Volkoff (TOV)
equation~\cite{Tolman1939,Oppenheimer1939} in the presence of charge is given by
\begin{equation}
-\frac{M_G(\rho+p_r)}{r^2}e^{\frac{\lambda-\nu}{2}}-\frac{dp}{dr}+
\sigma \frac{q}{r^2}e^{\frac{\lambda}{2}} =0, \label{eq25}
\end{equation}

where $M_G$ is the effective gravitational mass given by:

\begin{equation}
M_G(r)=\frac{1}{2}r^2 \nu^{\prime}e^{(\nu - \lambda)/2}.\label{eq26}
\end{equation}

Plugging the value of $M_G(r)$ in equation (\ref{eq25}), we get
\begin{equation}
-\frac{\nu'}{2}(\rho+p_r)-\frac{dp}{dr}+\sigma \frac{q}{r^2}e^{\frac{\lambda}{2}} =0,  \label{eq27}
\end{equation}

The above equation can be expressed into three different components gravitational $(F_g)$, hydrostatic $(F_h)$ and electric $(F_e)$, which are defined as:

\begin{equation}
\label{eq28}
F_g=-\frac{\nu'}{2}(\rho+p_r)=-\frac{n\,A^2\,r}{4\,\pi}\,\frac{\left[-D\,\psi^n\,(1+Ar^2)+ n\,\psi^2 + 2n\,D\,Ar^2\,\psi^n\right]}{[\psi^2+D\,Ar^2\,\psi^n]^2}
\end{equation}

\begin{equation}
F_h=-\frac{dp_r}{dr}  \label{eq29}
\end{equation}

\begin{equation}
 \label{eq30}
F_e=\frac{A^2\,r}{4\,\pi}\,\frac{\left[\,2 D n^3\, \psi^{n+1} A^2r^4 + n^2\,F_{e1} + D\,\psi^n \,F_{e2} - 2 n\,F_{e3}\,\right]}{2\,[\psi^2 + D\,Ar^2\,\psi^n]^3}
\end{equation}

where,\\
$F_{e1}=[3-(10+ D\,\psi^n)Ar^2 +2(6-2 D \,\psi^n + D^2\,\psi^{2 n})A^2r^4 - (6-5\,D\,\psi^{n}) A^3r^6 + A^4r^8]$,\\

$F_{e2}=[-6 \psi^2 + D^2 \psi^{2 n}\, Ar^2 + D \psi^n \,(3 - 4 Ar^2 + 3 A^2r^4)]$,\\

$F_{e3}=[-(Ar^2-3) \psi^3 + 2 D^2 \psi^{2 n}\, A^2r^4 + D\,\psi^n (-3 + 6 Ar^2 - 5 A^2r^4 + 2 A^3r^6)]$.

The balancing of these three forces within the stellar interior leads to hydrostatic equilibrium of the fluid sphere.

\begin{figure}[!htbp]\centering
    \includegraphics[width=6cm]{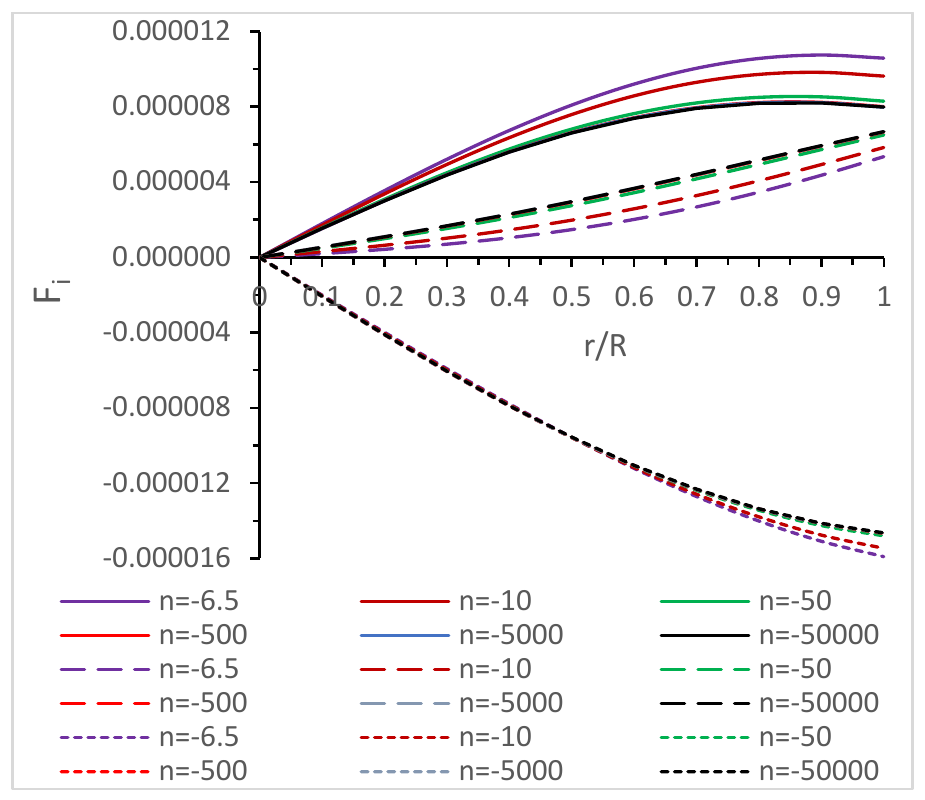}
\caption{Variation of different forces, $F_h$ (solid lines), $F_e$ (long dash lines), $F_g$ (dotted lines) with the radial coordinate ($r/R$) for $4U 1538-52$ with mass $(M)= 0.87 M_\odot$ and radius $(R)=7.866Km $ (Table \ref{TableA1}). For plotting of this figure, we have employed same values of the constant as used in Figs.6 and 7. The corresponding numerical values can be seen from Table\ref{TableA3}.}
    \label{Fig8}
\end{figure}

\subsubsection{Stability through adiabatic index}
The stability of the charged fluid models depends on the adiabatic index $\gamma$. Heintzmann and Hillebrandt \cite{Heintzmann1975} proposed that a neutron star model with equation of state is stable if $\gamma > 1$. This condition for stable model is necessary but not sufficient model (\cite{Tupper1983}). In the Newton's theory of gravitation, it is also well known that there has no upper mass limit if the equation of state has an adiabatic index $\gamma > 4/3$.

\begin{equation}
\Gamma=\frac{p+\rho}{p}\,\frac{dp}{d\rho}\, \label{eq31}
\end{equation}

Relation (\ref{eq31}) arises from an assumption within the Harrison-Wheeler formalism\cite{wheel1}. Chan et al. \cite{channy} in their study of dissipative gravitational collapse of an initially static matter distribution which is perturbed showed that Eq. (\ref{eq31}) follows from the equation of state of the unperturbed, static matter distribution.  In the case of anisotropic fluids the ratio of the specific heats assumes the following form
\begin{equation} \label{eqgamma}
\Gamma > \frac{4}{3} - \left[\frac{4}{3}\frac{p_r - p_t}{rp_r'}\right]_{max}\end{equation}
As pointed earlier a charged mass distribution can be viewed as an anisotropic system in which the radial and tangential stresses are unequal. In the case of isotropic pressure ($p_r = p_t$) we regain the classical Newtonian result from Eq. (\ref{eqgamma}). It is clear from Eq. (\ref{eqgamma}) that the instability is increased when $p_r < p_t$ and decreases when $p_r > p_t$.
\begin{figure}[!htbp]\centering
    \includegraphics[width=5.5cm]{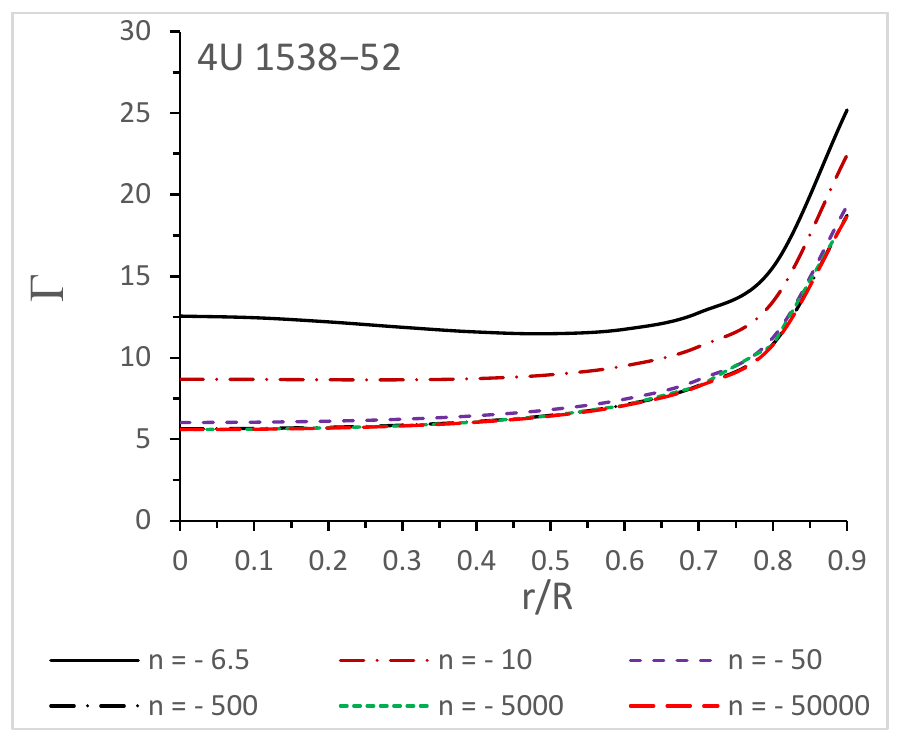}
\caption{Variation of adiabatic index $\Gamma$ with the radial coordinate ($r/R$) for $4U 1538-52$ with mass $(M)= 0.87 M_\odot$ and radius $(R)=7.866Km $ (Table \ref{TableA1}). For plotting of this figure, The values of constants for different $n$ are as follows: (i) A=5.3980 $\times 10^{-4} $,  B=0.548101, D=10.38552,  K=8.30822$\times 10^{2}$   for $n=-6.5$, (ii) A=3.5121$\times 10^{-4} $ , B=0.549818, D=16.26344, K=8.422215$\times 10^{2}$ for $n=-10$, (iii) A=7.0350$\times 10^{-5}$, B=0.552284, D=83.46073, K=8.592418 $\times 10^{2}$ for $n=-50$, (iv) A=7.03753$\times 10^{-6}$, B=0.552841, D=8.39474$\times 10^{2}$, K=8.630715 $\times 10^{2}$ for $n=-500$, (v) A=7.03753$\times 10^{-7}$, B=0.552899, D=8.39963$\times 10^{3}$, K=8.634831 $\times 10^{2}$ for $n=-5000$, (vi) A=7.03753$\times 10^{-8}$, B=0.552905, D=8.40010$\times 10^{4}$, K=8.635231 $\times 10^{2}$ for $n=-50000$ (Table \ref{TableA3}).}
    \label{Fig8}
\end{figure}

\subsubsection{Harrison-Zeldovich-Novikov static stability criterion}

In order for the configuration to be stable, the Harrison-Zeldovich-Novikov static stability criterion requires that the mass of the star increases with central density i.e. $dM/d\rho_0 > 0$ and unstable if $dM/d\rho_0 \le 0$.

\begin{equation}
M=\frac{R\,{\rho_1}^2\, [3\, D^2 \,\Phi^{2\,n}\, \rho_1 + n\,(n-2 )\,\Phi^2\,\rho_1] + 2\,D\,R\,\rho_1\,\Phi^{n+1}\,[\,1+(n-2)\,\rho_1\,]}{4\,[\Phi^{2} + D\,\rho_1\, \Phi^n]^2}
\end{equation}

\begin{equation}
\frac{dM}{d\rho_0}=\frac{R^3\,[\,M_1+M_2+M_3+M_4\,]}{2\,[\Phi^{2} + D\,\rho_1\, \Phi^n]^3}
\end{equation}

where $\rho_1=\frac{8\,\pi\,\rho_0}{3\,D}\,R^2$, \,\,  $\Phi=1-\rho_1$, $\rho_0=$\,central density, \\

\,\, $M_1=D\, \Phi^n - (n-2)\,{\rho_1}^4\, [n-D\,\Phi^n + D\, n^2\,\Phi^n]$,\\

$M_2=\rho_1\,[n^2 + n\,(-2 + D \Phi^n) + 2\, D\,\Phi^n\, (-2 + D \Phi^n)]$,\\

$M_3={\rho_1}^3\,[D\,n^3\, \Phi^n -D\, \Phi^n\, (-2 + D\, \Phi^n) -3\, n\, (2 + D\, \Phi^n) + n^2\, (3 - D\,\Phi^n + D^2\,\Phi^{2\,n})]$,\\

$M_4=-{\rho_1}^2\,[-3\, D\, \Phi^n + n^2\, (3 + D\,\Phi^n) - n\, (6 + D\, \Phi^n - 2\, D^2\, \Phi^{2\,n})]$.

Fig. 10 shows that  $dM/d\rho_0 > 0$ thus indicating that our model is stable. We further note that $dM/d\rho_0$ is independent of $n$ for low density stars. It is clear that $dM/d\rho_0$ decreases as $|n|$ increases for high density stars.

\begin{figure}[h!] \centering
	\includegraphics[width=5.5cm]{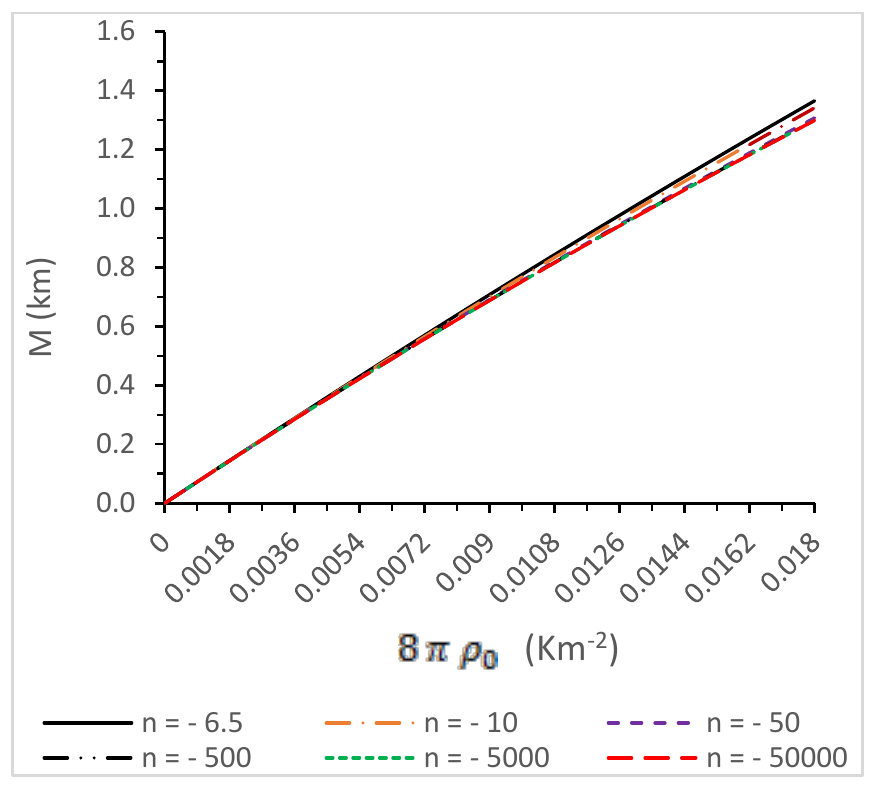}
	\caption{variation of Mass ($M$) versus central density $8\,\pi\,\rho_0 \,(0-9.666\times10^{14} gm/cm^3) $  for the anisotropic star 4U1608-52. For this graph we have employed numerical values for the constants same as used in Fig.9  (table 1).}
	\label{10}
\end{figure}

\subsection{Electric charge}

Table 1. displays the magnitude of the charge at the centre and
boundary for different stars. Also, from Fig.~3,
it is clear that the charge profile is zero at the centre (corresponding to vanishing electric field)
and monotonically increasing away from the centre, acquiring a maximum value at the boundary of the star. We further note that the charge increases with an increase in $|n|$ with the difference becoming indistinguishable at the stellar surface for very large $|n|$. We may then interpret $n$ as a 'stabilizing' factor. The variation of charge with $n$ suggests that lower values of $n$ imply lower charge which in turn means smaller electromagnetic repulsion. Fig.~3 shows that larger $|n|$ leads to greater surface charge thus indicating greater electromagnetic repulsion here. This would mean that the surface layers of the charged body is more stable than the inner core. The onset of collapse of such a body could proceed in an anisotropic manner or the collapse could lead to the cracking of the object thus avoiding the formation of a black hole.  As pointed out by Ray et al. \cite{rayb} the charge can be as high as $10^{20}$ coulombs and hydrostatic equilibrium may still be achieved however these equilibrium states are unstable. Bekenstein \cite{bek} argued that high charge densities will generate very intense electric fields. This will in turn induce pair production within the star thus destablizing the core. As an illustration we calculate the amount of charge at the boundary in coulomb unit for the compact star 4U1608-52 as follows: (i) $8.90468 \times10^{19}$ Coulomb for $n = - 6.5$, (ii) $9.52895\times10^{19}$ Coulomb for $ n = -10 $, (iii) $1.0370\times10^{20}$ Coulomb for $n = - 50$, (iv) $1.05471\times10^{20}$ Coulomb for $n= - 500$, (v) $1.05645\times10^{20}$ Coulomb for $n= - 5000$, (vi)  $1.05662\times10^{20}$ Coulomb for $n = - 50000$. However, the amount of charge in coulomb unit throughout the star can be determined by multiplying every recorded value in table 1. by a factor of $1.1659\times10^{20}$.

\begin{table}
\centering \caption{The electric charge for compact star 4U 1538-52 for different values of $n$ in the
relativistic unit (km).} \label{Table1}

{\begin{tabular}{@{}ccccccc@{}} \hline

$r/a$ & $n$ = - 6.5 & $n$ = - 10 & $n$ = - 50 & $n$ = - 500 & $n$ = - 5000 & $n$ = - 50000 \\ \hline

0.0 & 0& 0 &  0 & 0 & 0 & 0\\ \hline

0.2 &0.004136 & 0.005083 & 0.006337 & 0.006603 & 0.006630 & 0.006632 \\ \hline

0.4 &0.035663 & 0.042614 &  0.051869 & 0.053831 & 0.054026 & 0.054045 \\ \hline

0.6 &0.133453 & 0.153669 &  0.180723 & 0.186445 &0.187010 & 0.187067 \\ \hline

0.8 &0.353352 & 0.391594 &  0.442957 & 0.453784 & 0.454852 & 0.454958 \\ \hline

1.0 &0.763760 & 0.817304 & 0.889463 & 0.904631 & 0.906125 & 0.906273 \\ \hline

\end{tabular}}
\end{table}

\subsection{Effective mass and compactness parameter for the charged compact star}
The maximal absolute limit of mass-to-radius $(M/R)$ ratio as proposed by Buchdahl\cite{Buchdahl1959} for static
spherically symmetric isotropic fluid models is given by $2M/R\leq 8/9 $. On the other hand,~\cite{Boehmer2006} proved that for a compact charged fluid sphere there is a lower bound for the mass-radius ratio
\begin{equation}
\frac{Q^{2}\, (18 R^2+ Q^2) }{2R^{2}\, (12R^2+Q^2)}  \leq
\frac{M}{R}, \label{eq62}
\end{equation}
for the constraint $Q < M$.

However this upper bound of the mass-radius ratio for charged compact star was
generalized by~\cite{And} who proved
that
\begin{equation}
\frac{M}{R} \leq \left[\frac{4R^2+3Q^2}{9R^2} +\frac{2}{9R}\,\sqrt{R^2+3Q^2}\right]. \label{eq63}
\end{equation}

The  Eqs. \ref{eq62} and \ref{eq63} imply that
\begin{equation}
\frac{Q^{2}\, (18 R^2+ Q^2) }{2R^{2}\, (12R^2+Q^2)}  \leq
\frac{M}{R} \leq \left[\frac{4R^2+3Q^2}{9R^2} +\frac{2}{9R}\,\sqrt{R^2+3Q^2}\right]
\end{equation}

The effective mass of the charged fluid sphere can be determined as:
\begin{equation}
m_{eff}=4\pi{\int^R_0{\left(\rho+\frac{E^2}{8\,\pi}\right)\,r^2\,dr}}=\frac{R}{2}[1-e^{-\lambda(R)}]\, \label{eq32}
\end{equation}
where $e^{-\lambda}$ is given by the equation (\ref{eq10b})

and compactness $u(r)$ is defined as:

\begin{equation}
u(R)=\frac{m_{eff}(R)}{R}=\frac{1}{2}[1-e^{-\lambda(R)}]\, \label{eq33}
\end{equation}

\subsection{Redshift}

The maximum possible surface redshift for a bounded configuration with isotropic pressure is $Z_s = 4.77$. Bowers and Liang showed that this upper bound can be exceeded in the presence of pressure anisotropy\cite{bowers}. When the anisotropy parameter is positive ($p_t > p_r$) the surface redshift is greater than its isotropic counterpart. Haensel et al. \cite{strange} showed that for strange quark stars the surface redshift is higher in low mass stars with the difference being as high as 30$\%$ for a 0.5 solar mass star and 15$\%$ for a 1.4 solar mass star. The gravitational surface red-shift ($Z_s$) is given as:

\begin{equation}
Z_s= (1-2\,u)^{\frac{-1}{2}} -1=\sqrt{1+D\,AR^2\,(1-AR^2)}-1, \label{zs}
\end{equation}

From Eq.(\ref{zs}), we note that the surface redshift depends upon the compactness $u$, which implies that the surface redshift for any star can not be arbitrary large because compactness $u$ satisfies the Buchdhal maximal allowable mass-radius ratio. However, surface redshift will increase with increase of compactness $u$. Also, from Table 5. we observe that the surface redshift decreases with an increase in $|n|$.

\begin{figure}[!htbp]\centering
    \includegraphics[width=5.5cm]{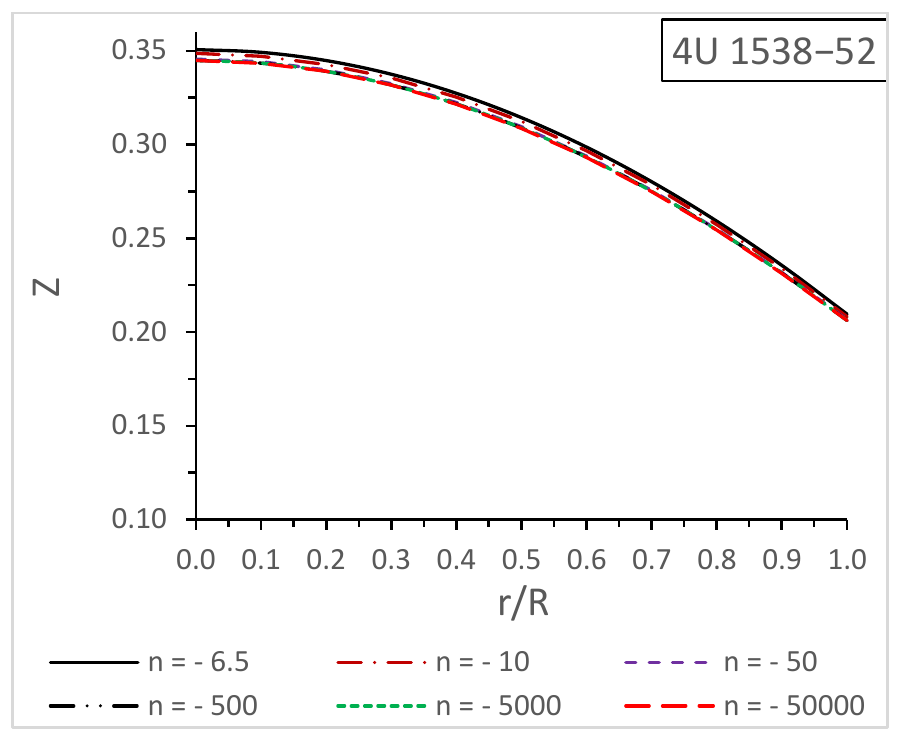}
\caption{Variation of redshift ($Z$) with the radial coordinate ($r/R$) for $4U 1538-52$ with mass $(M)= 0.87 M_\odot$ and radius $(R)=7.866Km $ (Table \ref{TableA1}). For plotting of this figure, The values of constants for different $n$ are as follows: (i) A=5.3980 $\times 10^{-4} $,  B=0.548101, D=10.38552,  K=8.30822$\times 10^{2}$   for $n=-6.5$, (ii) A=3.5121$\times 10^{-4} $ , B=0.549818, D=16.26344, K=8.422215$\times 10^{2}$ for $n=-10$, (iii) A=7.0350$\times 10^{-5}$, B=0.552284, D=83.46073, K=8.592418 $\times 10^{2}$ for $n=-50$, (iv) A=7.03753$\times 10^{-6}$, B=0.552841, D=8.39474$\times 10^{2}$, K=8.630715 $\times 10^{2}$ for $n=-500$, (v) A=7.03753$\times 10^{-7}$, B=0.552899, D=8.39963$\times 10^{3}$, K=8.634831 $\times 10^{2}$ for $n=-5000$, (vi) A=7.03753$\times 10^{-8}$, B=0.552905, D=8.40010$\times 10^{4}$, K=8.635231 $\times 10^{2}$ for $n=-50000$ (Table \ref{TableA3}).}
    \label{Fig8}
\end{figure}

\begin{figure}[!htbp]\centering
    \includegraphics[width=5.5cm]{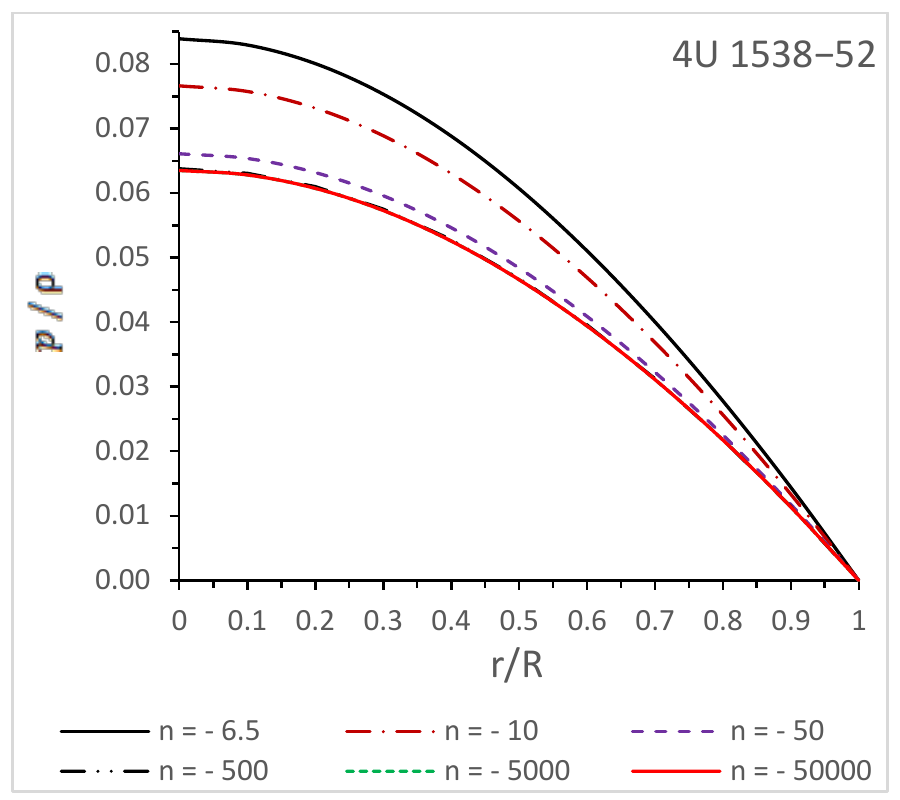}
\caption{Variation of the ratio $\frac{p}{\rho}$ with respect to the radial coordinate ($r/R$) for $4U 1538-52$ with mass $(M)= 0.87 M_\odot$ and radius $(R)=7.866Km $ (Table \ref{TableA1}). For plotting of this figure, The values of constants for different $n$ are as follows: (i) A=5.3980 $\times 10^{-4} $,  B=0.548101, D=10.38552,  K=8.30822$\times 10^{2}$   for $n=-6.5$, (ii) A=3.5121$\times 10^{-4} $ , B=0.549818, D=16.26344, K=8.422215$\times 10^{2}$ for $n=-10$, (iii) A=7.0350$\times 10^{-5}$, B=0.552284, D=83.46073, K=8.592418 $\times 10^{2}$ for $n=-50$, (iv) A=7.03753$\times 10^{-6}$, B=0.552841, D=8.39474$\times 10^{2}$, K=8.630715 $\times 10^{2}$ for $n=-500$, (v) A=7.03753$\times 10^{-7}$, B=0.552899, D=8.39963$\times 10^{3}$, K=8.634831 $\times 10^{2}$ for $n=-5000$, (vi) A=7.03753$\times 10^{-8}$, B=0.552905, D=8.40010$\times 10^{4}$, K=8.635231 $\times 10^{2}$ for $n=-50000$ (Table \ref{TableA3}).}
    \label{Fig9}
\end{figure}

\subsection{Equation of state}
An equation of state (EoS), $p = p(\rho)$ relates the pressure and the density of the stellar fluid and is an important indicator of the nature of the matter making up the configuration. The MIT bag model arising from observations in fundamental particle physics relates the pressure to the density of the star via a linear relation of the form $p = \alpha \rho - B$ where $B$ is the Bag constant. This equation of state has been successfully used to model compact objects in general relativity ranging from neutron stars through to strange star candidates. A recent model of a radiating star in which the collapse proceeds from an initial static configuration obeying a linear equation of state of the form $p_r = \alpha (\rho - \rho_s)$ where $p_r$ is the radial pressure, $\rho_s$ is the surface energy density and $\alpha$ is the EoS parameter showed that the variation of $\alpha$ affects the temperature profile of the collapsing body. Fig. 9 shows the variation of the ratio $p/\rho$ with $r/R$. We note that the pressure is less than the density at each interior point of the configuration. This ratio is also positive everywhere inside the star. As $|n|$ increases, the ratio $p/\rho$ decreases with the differences tending to zero towards the surface layers of the star.

\begin{table}
	\centering
	\caption{Comparison between estimated and observed values of mass and radius for different compact stars \cite{GA}}
	\label{TableA1}
	\begin{tabular}{@{}lrrrrrrr@{}} \hline
		Compact star & $M/M_{\odot}$ & $R\,(Km)$ & $M/R$ & $M/M_{\odot}$& $R\,(Km)$  \\
          &$(estimated)$ & $(estimated)$ & $(estimated)$ & $(observed)$ & $(observed)$  \\\hline
		 4U 1538-52 & 0.87 & 7.866 & 0.162938 & 0.87 $\pm$ 0.07 & 7.866  $\pm$ 0.21   \\
	     SAX J1808.4-3658 & 0.90 & 7.951  & 0.166756 & 0.9 $\pm$ 0.3 & 7.951 $\pm$ 1.0  \\ \hline
		
	\end{tabular}
\end{table}

\begin{table}
\centering \caption{Numerical data of $AR^2$ corresponding to observed mass and radius with reference to table 1 for different values of $n$ }
 \label{TableA2}

{\begin{tabular}{@{}ccccccccc@{}}

\hline &  $ n=-6.5$ & $n=-10$ & $n=-50$ & $n=-500$ &$n=-5000$ &$n=-50000$\\

\hline Compact stars & $A{R}^{2}$ & $A{R}^{2}$ & $A{R}^{2}$ & $A{R}^{2}$ & $A{R}^{2}$ & $A{R}^{2}$\\

\hline  4U 1538-52 & 0.02117 & 0.008472 & 0.0016948 & 0.00016949 & 0.000016949 & 0.0000016949 \\

         SAX J1808.4-3658  & 0.034453 & 0.02242 & 0.0044913 & 0.0004493 & 0.00004493 & 0.000004493 \\

\hline
\end{tabular}}
\end{table}

\begin{table}
\centering \caption{Numerical data of physical parameters
 $AR^2$, $A$, $B$,  $D$ ,  $K$  and $nA$ for the different
values of $n$ for $4U 1538-52$}\label{TableA3}
{\begin{tabular}{@{}ccccccc@{}}

\hline

$n$  & $AR^2$ & $A(km^{-2})$ & $B$ & $D$ & $K(km^2)$  & $nA$ \\ \hline

$-6.5$ & 0.033400 &   5.3980 $\times 10^{-4} $  &  0.548101 & 10.38552 &  8.30822$\times 10^{2} $ & -0.0035087 \\
$-10$ & 0.021731 &  3.5121$\times 10^{-4} $   & 0.549818 & 16.26344 & 8.422215$\times 10^{2} $ &  -0.0035121 \\
$-50$ & 0.004353 & 7.0350$\times 10^{-5}$  &  0.552284 & 83.46073 & 8.592418 $\times 10^{2} $ & -0.0035175 \\
$-500$ & 0.00043544 & 7.03753$\times 10^{-6}$  & 0.552841 & 8.39474$\times 10^{2}$ & 8.630715 $\times 10^{2} $ & -0.003518765 \\
$-5000$ & 0.000043545 &  7.03753$\times 10^{-7}$ & 0.552899 & 8.39963$\times 10^{3}$ & 8.634831 $\times 10^{2}$ &  -0.003518765 \\
$-50000$ & 0.0000043545 & 7.03753$\times 10^{-8}$& 0.552905 & 8.40010$\times 10^{4}$  & 8.635231 $\times 10^{2}$ & -0.003518765 \\ \hline
\end{tabular}}
\end{table}

\begin{table}
\centering \caption{Numerical data of physical parameters
 $AR^2$, $A$,$B$, $D$ ,  $K$  and $nA$ for the different
values of $n$ for $SAX J1808.4-3658$}\label{TableA4}
{\begin{tabular}{@{}ccccccc@{}}

\hline

$n$  & $AR^2$ & $A(km^{-2})$ & $B$ & $D$ & $K(km^2)$  & $nA$ \\ \hline

$-6.5$ &0.034453 & 5.4496$\times 10^{-4} $  &  0.538534 & 10.3087854&  8.313872$\times 10^{2} $ & -0.00354224 \\
$-10$ & 0.02242 &3.5464$\times 10^{-4} $   & 0.540283& 16.1520157 &8.42981$\times 10^{2} $ &  -0.0035464 \\
$-50$ & 0.0044913 & 7.1039$\times 10^{-5}$  & 0.552284 &83.4607306 & 8.509081 $\times 10^{2} $ & -0.00355195 \\
$-500$ & 0.0004493 & 7.1070$\times 10^{-6}$  & 0.543421 &  8.345640$\times 10^{2}$ & 8.643643 $\times 10^{2} $ & -0.0035535  \\
$-5000$ & 0.00004493 & 7.1070$\times 10^{-7}$ & 0.5434891 &8.350693$\times 10^{3}$ & 8.647794 $\times 10^{2}$ & -0.0035535 \\
$-50000$ &0.000004493& 7.1070$\times 10^{-8}$&  0.5434959 & 8.351198$\times 10^{4}$  & 8.648210 $\times 10^{2}$ & -0.0035535 \\ \hline
\end{tabular}}
\end{table}

\begin{table}
	\centering
	\caption{The central density, surface density and central pressure for compact star candidate $4U 1538-52$ }\label{TableA5}
	
\begin{tabular}{ccccc} \hline
value & Central Density & Surface Density & Central Pressure & Surface  \\
of $n$ & $(gm/cm^{3}) $ & $(gm/cm^{3})$ & $(dyne/cm^{2})$ & Redshift \\\hline
$-6.5$  & 9.0314$\times 10^{14} $ & 7.52234$\times 10^{14} $ & 6.82219$\times 10^{34}$ &$0.20954368$   \\
$-10$ &9.2018$\times 10^{14} $ & 7.38531$\times 10^{14} $ & 6.34375$\times 10^{34}$ & $0.208319814$  \\
$-50$ & 9.4589$\times 10^{14} $ &7.18798$\times 10^{14} $ & 5.62454$\times 10^{34}$ & $0.206572451$  \\
$-500$ &  9.5175$\times 10^{14} $ &  7.1447$\times 10^{14} $ & 5.4610$\times 10^{34}$ & $0.206180642$   \\
$-5000$ & 9.5230$\times 10^{14} $ & 7.14015$\times 10^{14} $ & 5.4444$\times 10^{34}$ & $0.206139815$  \\
$-50000$ & 9.5236$\times 10^{14} $ &7.1397$\times 10^{14} $ & 5.4427$\times 10^{34}$ & $0.206135287$   \\ \hline
\end{tabular}
\end{table}

\section{Discussion}

In this paper we attempted to obtain electromagnetic mass models (EMMM) which were first addressed by Lorentz. The Lorentz electromagnetic mass models had the distinguishing feature that vanishing charge density is accompanied by the simultaneous vanishing of all other thermodynamical quantities. In addition, the equation of state of these models is of the form $\rho + p = 0$ giving rise to negative pressure. The solution obtained in this work relaxes this particular equation of state, allowing for positive pressure. The gravitational and thermodynamical behaviour of our model is controlled by a parameter $n$. Switching off $n$ results in the vanishing of charge density and all other thermodynamical quantities such as density and pressure. We use a novel approach of embedding a spherically symmetric, static  metric in Schwarzschild coordinates into a five-dimensional flat metric. This embedding is equivalent to the Karmarkar condition: the requirement for a spherically symmetric metric to be of embedding class 1. The condition obtained from this embedding relates the gravitational potentials thus reducing the problem of finding an exact solution of the Einstein-Maxwell field equations to a single-generating function. By specifying one of the gravitational potentials on physical grounds, we obtain the second potential which completely describes the gravitational behaviour of the compact object. The junction conditions required for the smooth matching of the interior spacetime to the exterior Reissner-No$\ddot{r}$dstrom spacetime fixes the constants in our solution and determines the mass contained within the charged sphere. Our model displays many salient features which are bodes well for describing a compact, self-gravitating object. Graphical analysis of the solution shows that the density and pressure are monotonically decreasing functions of the radial coordinate. The pressure vanishes at some finite radius. This indicates that our solution can be utilised to describe a bounded object unlike the Kohler-Chao solution which arises from imposition of the Karmarkar condition together with pressure isotropy. Causality is obeyed at each interior point of the configuration. Stability analysis via the adiabatic index and the Harrison-Zeldovich-Novikov static stability criterion indicate that our model is stable. Analysis of the variation of charge with the radial coordinate reveals an interesting characteristic of our model. The charge increases with the parameter $|n|$. This increase is largest towards the surface layers of the charged object becoming simultaneously indistinguishable for very large values at the surface. This implies that the surface layers are more stable (larger repulsive forces here) than inner core layers. This 'differentiated' stability may lead to anisotropic collapse or the subsequent cracking of the sphere should this object starts to collapse. This phenomenon has not been discussed elsewhere in the literature. The influence of the parameter $n$ is clearly drawn out in tables 1 - 6. Table 2. shows that our theoretical model describes compact objects to a very good degree of accuracy with regards to observed masses and radii of stars. Tables 3 to 5 clearly show that variations in the model parameters stabilise for very large $n$. Table 6. illustrates the influence of the parameter $n$ on the central density, surface density, central pressure and surface redshift. It is clear that for very large $n$ variations in these physical quantities tend to zero. This feature indicates that the parameter $n$ can be viewed as a 'building' constant, that is to say, that an increase in $n$ is accompanied by an increase in mass, radius and charge which builds up the star from $r = 0$ through to the surface. In this work we have utilised $n < 0$ and the case $n \geq 0$ was studied by \cite{maurya11}. Future work has been initiated to consider the case of general $n$.

{ }

\end{document}